\documentclass[reprint, amsmath, amssymb, aps, prb, superscriptaddress, citeautoscript]{revtex4-1}

\usepackage{graphicx}% Include figure files
\usepackage{bm}
\usepackage{array}
\newcolumntype {s}[1]{@{\hspace{#1}}} % space
\newcolumntype {R}{>{$}r<{$}}         % switch text <-> math
\newcolumntype {C}{>{$}c<{$}}         % switch text <-> math
\newcolumntype {L}{>{$}l<{$}}         % switch text <-> math
\newcolumntype {f}{@{\extracolsep\fill}}  % fill
\usepackage{dcolumn}% Align table columns on decimal point
\newcommand* {\tvek}[2][c]{\left( \begin{array}{s{0.15em}#1s{0.15em}}
     #2\end{array} \right)}

\let\Im\relax\DeclareMathOperator{\Im}{Im}
\newcommand* {\vek}[1]{{\bm{\mathrm{#1}}}}
\newcommand* {\vekc}[1]{{\bm{\mathcal{#1}}}}
\newcommand* {\kk}{\vek{k}}

\newcommand* {\frack}[2]{{\textstyle\frac{#1}{#2}}}
\newcommand* {\braket}[1]{\langle {#1} \rangle}
\newcommand* {\pseudo}{\mbox{(pseudo-)}}

\begin{document}
\title{Universal spin dynamics in quantum wires}
\author{E. A. Fajardo}
\email{eafajardo@niu.edu}
\altaffiliation[On leave from: ]{Department of Physics, Mindanao State University-Main Campus, Marawi City, Lanao del Sur, Philippines 9700}
\affiliation{Department of Physics, Northern Illinois University,
DeKalb, IL 60115, USA}

\author{U. Z\"ulicke}
\affiliation{School of Chemical and Physical Sciences and
MacDiarmid Institute for Advanced Materials and Nanotechnology,
Victoria University of Wellington, PO Box 600, Wellington 6140,
New Zealand}

\author{R. Winkler}
\affiliation{Department of Physics, Northern Illinois University,
DeKalb, IL 60115, USA}
\affiliation{Materials Science Division, Argonne National
Laboratory, Argonne, IL 60439, USA}

\date{\today}

\begin{abstract}
  We discuss the universal spin dynamics in quasi one-dimensional
  systems including the real spin in narrow-gap semiconductors like
  InAs and InSb, the valley pseudospin in staggered single-layer
  graphene, and the combination of real spin and valley pseudospin
  characterizing single-layer transition metal dichalcogenides
  (TMDCs) such as MoS$_2$, WS$_2$, MoSe$_2$, and WSe$_2$.  All these
  systems can be described by the same Dirac-like Hamiltonian.
  Spin-dependent observable effects in one of these systems thus
  have counterparts in each of the other systems.  Effects discussed
  in more detail include equilibrium spin currents, current-induced
  spin polarization (Edelstein effect), and spin currents generated
  via adiabatic spin pumping.  Our work also suggests that a
  long-debated spin-dependent correction to the position operator in
  single-band models should be absent.
\end{abstract}

\maketitle

\section{Introduction}

Spintronics seeks to exploit the spin degree of freedom in order to
achieve new or more efficient functionalities not available in
charge-based electronics \cite{wol01, zut04, die08a}.  The
spin degree of freedom emerges from a relativistic treatment of the
electrons' motion.  The electrons' spin interacts with the
orbital environment via spin-orbit (SO) coupling that scales with the
atomic number $Z$ of the atoms involved, so that it is advantageous
to use high-$Z$ materials.  On the other hand, spin-orbit coupling
in novel 2D materials such as graphene with $Z=6$ is found to be
negligible \cite{net09}.

In many materials the electrons near the Fermi energy reside in
multiple inequivalent valleys, which give rise to yet another degree
of freedom called the valley pseudospin \cite{shk02, gun06, ryc07,
xia07, xia12}, and valleytronics seeks to exploit the valley
pseudospin for new device functionalities \cite{ sch16}.  A major
advantage of valleytronics over spintronics lies in the fact that it
allows one to reach parameter regimes not available for the real
spin in, e.g., low-$Z$ materials such as graphene.

Previous work has touched on the conceptual similarities between, on
the one hand, the real spin and spin-orbit coupling and, on the
other hand, the valley pseudospin and valleyspin-orbit coupling
\cite{ryc07, xia07, xia12, sch16}.  In the present paper we discuss
the \emph{universal} spin dynamics characterizing a diverse range of systems
in reduced dimensions, starting from the familiar fully relativistic
Dirac equation and a (simplified) Kane model \cite{kan57, win03}
suited for narrow-gap, high-$Z$ semiconductors such as InAs and
InSb, to multivalley systems that include staggered
single-layer graphene \cite{xia07} and transition metal
dichalcogenides \cite{xia12, kor15} (TMDCs) such as MoS$_2$, WS$_2$,
MoSe$_2$ and WSe$_2$.  Indeed, all these systems are characterized by
the same generic effective $4\times 4$ Hamiltonian, indicating that
we get analogous manifestations of real-spin and valley-pseudospin
dynamics.  For concreteness, we focus on examples in quasi-1D quantum
wires.

This paper is organized as follows. In Sec.~\ref{sec:eff-ham}, we
discuss the formulation of the generic effective Hamiltonian,
where for different physical systems represented by this Hamiltonian
the spin operator matches the real spin, the valley spin or an
entangled combination of both the real spin and valley
pseudospin. In Sec.~\ref{sec:quanwire} we apply this model to quasi-1D quantum wires, focusing on a range of problems.  It has
long been debated \cite{yaf63, noz73, bae90, eng07a, bi13} whether the position
operator in a multiband system, when projected on the subspace of
positive or negative energies, should acquire a spin-dependent
correction that manifests itself as a factor of two for a
spin-dependent correction for the velocity operator.  Our study
suggests that the spin-dependent correction for the position
operator and the factor of two in the velocity operator should be
absent.  We use these results for the velocity operator to discuss
equilibrium spin currents in quantum wires \cite{ras03, mal03,
kis05}.  Furthermore, we discuss the Edelstein effect \cite{ede90,
aron91} for quantum wires, where an electric field driving a
dissipative charge current gives rise to a \pseudo\ spin
polarization.  Finally, we discuss adiabatic \pseudo\ spin pumping
\cite{bro98, gov03} as a means to generate a \pseudo\ spin current.
Section~\ref{sec:conclusion} presents the conclusions.

\section{Effective 2D Hamiltonian}
\label{sec:eff-ham}

We consider the generic effective $4\times 4$ Hamiltonian in 2D
\begin{subequations}
  \label{eq:ham:4x4}
  \begin{align}
    H_{4\times 4} & = H_0 + H_1, \\
    H_0 & = \frack{1}{2} \Delta \, \sigma_0 \rho_z, \\
    H_1 & = \gamma \left(k_x \sigma_x \rho_x + k_y \sigma_y \rho_x \right)
    + e\vekc{E}\cdot\vek{r} 
  \end{align}
\end{subequations}
for the motion in a 2D plane.  Here $\rho_i$ and $\sigma_i$ denote
Pauli matrices, we have $\sigma_0 = \openone_{2\times 2}$, $\Delta$ is
the energy gap, $\hbar\kk = -i\hbar\nabla + e\vek{A}$ is the
operator of (crystal) kinetic momentum, and $e\vekc{E}\cdot\vek{r}$ is the
potential due to an electric field $\vekc{E}$.  More explicitly, we
have
\begin{subequations}
\begin{equation}
  H_{4\times 4} = \tvek[cccc]{
  \frac{\Delta}{2} & 0 & 0 & \gamma k_- \\
  0 & \frac{\Delta}{2} & \gamma k_+ & 0 \\
  0 & \gamma k_- & - \frac{\Delta}{2} & 0 \\
  \gamma k_+ & 0 & 0 & - \frac{\Delta}{2}}  + e\vekc{E}\cdot\vek{r}_{4\times 4}
\end{equation}
with $k_\pm \equiv k_x \pm ik_y$ and $\vek{r}_{n\times n} \equiv \vek{r} \, \openone_{n\times n}$,
which is unitarily equivalent to
\begin{equation}
  \tilde{H}_{4\times 4} = \tvek[cc]{H_{2\times 2} & 0 \\ 0 & H_{2\times 2}^\ast}
\end{equation}
\end{subequations}
with
\begin{equation}
  H_{2\times 2} = \frack{\Delta}{2} \rho_z
  + \gamma \left(k_x \rho_x + k_y \rho_y \right)
    + e\vekc{E}\cdot\vek{r}_{2\times 2} ,
\end{equation}
indicating that the \pseudo\ spin up and down eigenstates of
$\sigma_z$ are completely decoupled.

The two-band Hamiltonian (\ref{eq:ham:4x4}) applies to a range of
systems.  In all examples discussed in the following, the Pauli
matrices $\rho_i$ define the subspaces of positive and negative
energies and their off-diagonal couplings, whereas the matrices
$\sigma_i$ represent a spin or pseudospin degree of freedom acting
within these bands.  The first realization of $H_{4\times 4}$ is the
Dirac Hamiltonian for systems confined to a 2D plane, where
$\Delta = 2mc^2$ and $\gamma = \hbar c$.  In this case, the matrices
$\sigma_i$ represent the real spin.

Second, $H_{4\times 4}$ represents a simple version of the Kane
Hamiltonian \cite{kan57, win03} for semiconductor systems in reduced
dimensions, where the upper (lower) band in Eq.\ (\ref{eq:ham:4x4})
characterized via the matrices $\rho_i$ becomes the conduction
(valence) band separated by the fundamental gap $\Delta$, while
$\sigma_i$ represents the real spin and $\gamma$ becomes Kane's
momentum matrix element (apart from a prefactor $\sqrt{2/3}$).  This
model is particularly suited for electron systems in narrow-gap
materials like InAs and InSb.

Third, the same Hamiltonian applies to staggered single-layer graphene,
where $\gamma/\hbar$ becomes the Fermi velocity and $\Delta$
characterizes the sublattice staggering \cite{xia07}. Just like in
the Kane model, the Pauli matrices $\rho_i$ characterize the
conduction and valence bands, yet $\sigma_i$ represents the valley
pseudospin.  The real spin and spin-orbit coupling can often be
ignored in graphene \cite{net09}.

\begin{figure}
   \includegraphics[width=0.95\columnwidth]{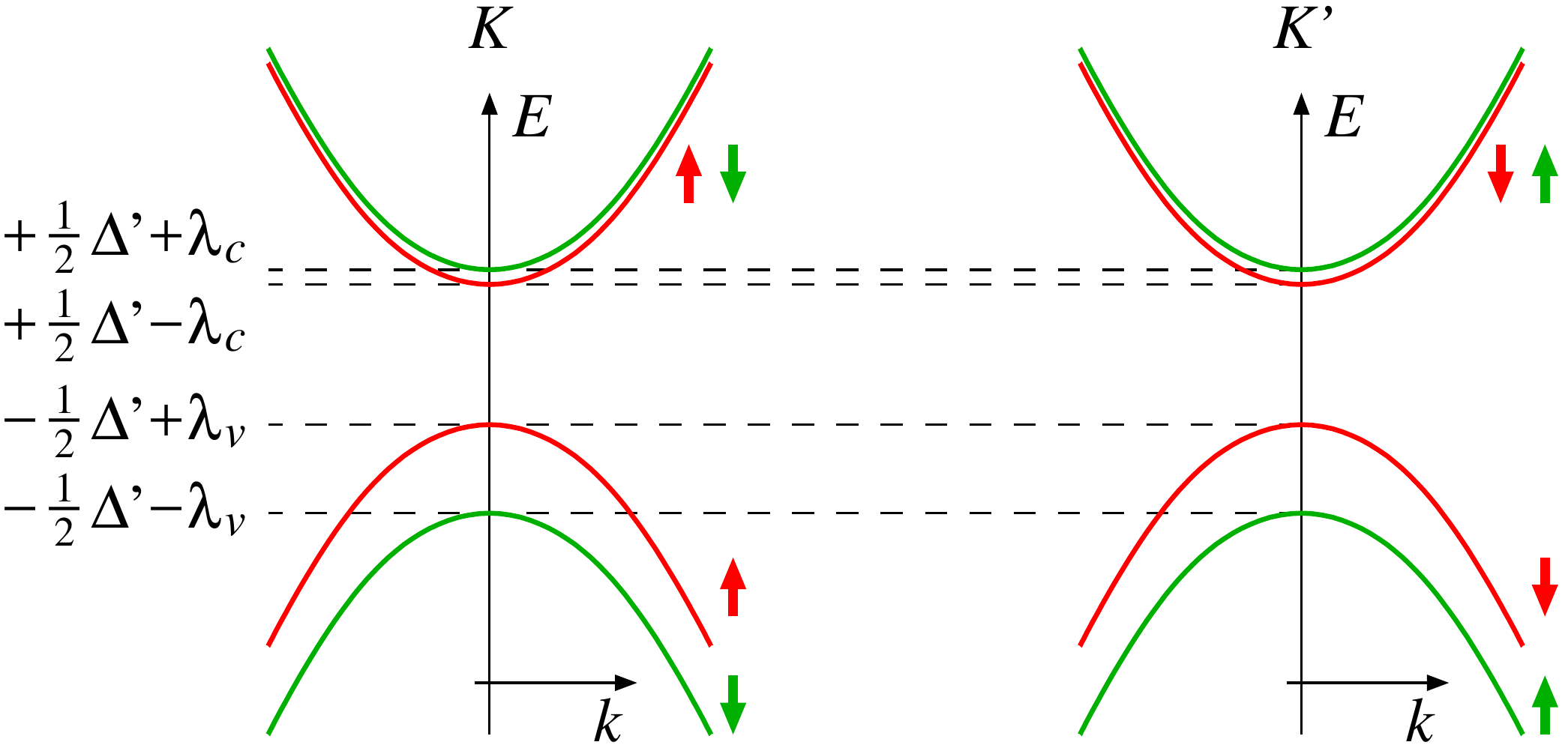}
  \caption{Energy dispersion near the points $K$ and $K'$ for the lowest conduction and highest valence band in MoS$_2$.  Here $\Delta'$ is the fundamental gap in the absence of SO coupling and $\lambda_c$ ($\lambda_v$) is the spin splitting in the conduction (valence) band.  The bands marked in red and green thus correspond to decoupled replicas of Hamiltonian (\ref{eq:ham:4x4}) with gaps
$\Delta^\pm = \Delta' \pm (\lambda_v + \lambda_c)$. 
It was found in Ref.~\onlinecite{kor15} that for WS$_2$ the sign of $\lambda_c$ is opposite to the sign in MoS$_2$ so that the ordering of the spin-split conduction bands near $K$ and $K'$ relative to the ordering of the spin-split valence bands is opposite to the one shown here.}\label{fig:valleyband}
\end{figure}

Lastly, the model (\ref{eq:ham:4x4}) can be applied to single layers
of TMDCs \cite{xia12, kor15} such
as MoS$_2$ and WS$_2$. Unlike graphene, the larger spin-orbit
coupling due to the transition metals' high atomic number gives rise
to a significant valley-dependent spin splitting in the valence and
(to a lesser extent) in the conduction band.  For each band
$i = v, c$, these splittings are of the form
$\lambda_i \Sigma^{(v)}_z \Sigma^{(s)}_z$, where $\Sigma^{(v)}_z$
($\Sigma^{(s)}_z$) is a Pauli matrix acting in valley pseudospin
(real spin) space.  The resulting band structure is depicted in
Fig.~\ref{fig:valleyband}.  The system can thus be described by two
decoupled replicas $H_{4\times 4}^\pm$ of the Hamiltonian
(\ref{eq:ham:4x4}) (corresponding to the red and green lines in
Fig.~\ref{fig:valleyband} and apart from a constant energy shift
between the replicas) with gaps
$\Delta^\pm = \Delta' \pm (\lambda_v + \lambda_c)$, where $\Delta'$
denotes the fundamental gap in the absence of SO coupling.  In this
case, the Pauli matrices $\sigma_i$ in Eq.\ (\ref{eq:ham:4x4})
represent an entangled combination of real spin and valley
pseudospin.  Depending on the position of the Fermi energy relative
to the bands in Fig.~\ref{fig:valleyband}, a complete description of
TMDCs via Eq.\ (\ref{eq:ham:4x4}) must take into account one or both
replicas $H_{4\times 4}^\pm$.  For brevity, we drop in the following
the superscript $\pm$, assuming $\Delta = \Delta^+$ or $\Delta^-$.

The material-dependent parameters $\Delta$ and $\gamma$ for the
various systems are listed in Table~\ref{table:parameter}.  The
numeric values are approximate.  The main purpose of this table is
to illustrate the range of numeric values of the parameters
$\Delta$ and $\gamma$ characterizing the different systems described by the
Hamiltonian (\ref{eq:ham:4x4}).  Evidently, the generic Hamiltonian
$H_{4\times 4}$ provides a unified treatment of the physics in these
systems, despite the rather different numeric values of the model
parameters and the different meanings of the \pseudo\ spin $\vek{\sigma}$.  Quite generally, any observable physics emerging from the Hamiltonian (\ref{eq:ham:4x4}) for one of these systems has a counterpart in the other systems.

\begin{table}
\caption{System parameters for the different realizations of the Hamiltonian (\ref{eq:ham:4x4}).  Numeric values are approximate.} \label{table:parameter}
\renewcommand{\arraystretch}{1.1}
\begin{tabular*}{\columnwidth}{lfcccc}
\hline \hline \rule{0pt}{2.5ex}
& $\Delta$ (eV) & $\gamma$ (eV\AA) & $\mu$ (eV\AA$^2$) & $-\alpha/\mathcal{E}_y$ ($e$\AA$^2$) \\
\hline
Dirac\rule{0pt}{2.5ex}
         & $1.0 \times 10^6$ & $2.0 \times 10^3$ & 7.6 & $3.7 \times 10^{-6}$ \\
InAs \footnote{Ref.~\onlinecite{win03}}     & 0.42 & 8.4 & $3.3 \times 10^2$ & $4.0 \times 10^2$ \\
InSb\footnotemark[1]    & 0.24 & 8.1 & $5.5 \times 10^2$ & $1.1 \times 10^3$ \\
graphene\footnote{Refs.~\onlinecite{xia07, net09}} & $\sim 0.1$ & 6.6 & $8.7 \times 10^2$ & $4.3 \times 10^3$ \\
MoS$_2$\footnote{Ref.~\onlinecite{xia12} }  & 1.7 & 3.5 & $1.4 \times 10^1$ & 4.2 \\
WS$_2$\footnotemark[3]   & 1.8 & 4.4 & $2.2 \times 10^1$ & 6.0 \\
MoSe$_2$\footnotemark[3] & 1.5 & 3.1 & $1.3 \times 10^1$ & 4.3 \\
WSe$_2$\footnotemark[3]  & 1.6 & 3.9 & $1.9 \times 10^1$ & 5.9 \\
\hline \hline
\end{tabular*}
\end{table}

In the Hamiltonian $H_{4\times 4} = H_0 + H_1$ the dynamics of the
subspaces of positive and negative energies are coupled via the
off-diagonal terms in $H_{4\times 4}$.  The Foldy-Wouthuysen (FW)
transformation \cite{fol50} (see also Refs.~\onlinecite{bir74,
win03}) is a unitary transformation $e^{-S}$ constructed by
successive approximations for the anti-Hermitian operator
$S = -S^\dagger$ such that
$H_{4\times 4}^\mathrm{FW} \equiv e^{-S} H_{4\times 4} \, e^S$
becomes block-diagonal.  This procedure, which is also known as
quasidegenerate perturbation theory, relies on the fact that we may
treat $H_0$ as unperturbed Hamiltonian and $H_1$ as perturbation.
In third order, this yields the block-diagonal Hamiltonian
\begin{equation}
  \label{eq:ham:4x4:mod}
  H_{4\times 4}^\mathrm{FW}
  \equiv e^{-S} H_{4\times 4} \, e^S
  = \tvek[cc]{\mathcal{H}_+ & 0 \\ 0 & \mathcal{H}_-}
\end{equation}
with effective $2\times 2$ Hamiltonians
\begin{align}
  \label{eq:ham:2x2}
  \mathcal{H}_\pm = &
    \left(\pm \frack{\Delta}{2} + e\vekc{E}\cdot\vek{r}\right)  \sigma_0
  + \frac{\gamma^2}{\Delta}(\pm k^2 \sigma_0 + \frack{e}{\hbar}B_z \,\sigma_z)
  \nonumber\\ &
  \mp \frac{e \, \gamma^2}{\Delta^2} (k_x \mathcal{E}_y - k_y \mathcal{E}_x) \, \sigma_z ,
\end{align}
where we used $[r_i, k_j] = i \delta_{ij}$ and
$[k_x,k_y]= - \frack{ie}{\hbar}B_z$.  By definition of the FW
transformation, the subspace of positive energies in Eq.\
(\ref{eq:ham:4x4:mod}) characterized by $\mathcal{H}_+$ is decoupled
from the subspace of negative energy characterized by
$\mathcal{H}_-$ so that $\mathcal{H}_+$ and $\mathcal{H}_-$ can be
discussed separately.  We also note that the problem exhibits
electron-hole symmetry, yet for definiteness we will focus in the
following on the subspace with positive energies.

\section{Effective Hamiltonian for quasi-1D Quantum Wire}
\label{sec:quanwire}

We illustrate the universal dynamics characterizing the different realizations of the Hamiltonians (\ref{eq:ham:4x4}) and (\ref{eq:ham:2x2}) by considering a quasi 1D wire along the $x$ direction.
Here, ignoring the quantized motion perpendicular to the wire and restricting ourselves to $B_z = 0$, the
effective Hamiltonian (\ref{eq:ham:2x2}) becomes
\begin{equation}\label{eq:ham:2x2:1d}
\mathcal{H} = \frack{1}{2} \mu k^2 + \alpha k\sigma_z ,
\end{equation}
where $k$ is the wave vector along the direction of the wire (dropping the subscript $x$).  From Eq.\ (\ref{eq:ham:2x2}), we have $\mu = 2 \gamma^2 / \Delta$.  The second term in Eq.\ (\ref{eq:ham:2x2:1d}) is a Rashba-type spin splitting \cite{byc84a} with $\alpha =  - (e\gamma^2 / \Delta^2) \mathcal{E}_y$ proportional to the electric field $\mathcal{E}_y$ perpendicular to the wire, which is tunable via external gates \cite{nit97}.

For later reference, we summarize some basic properties of quasi-1D electron
systems described by the effective Hamiltonian (\ref{eq:ham:2x2:1d}).  The
spin-dependent dispersion becomes
\begin{equation}\label{eq:dispersion}
E_\lambda(k) = \frack{1}{2}\mu k^2 + \lambda \alpha k ,
\end{equation}
where $\lambda=\pm$ characterizes the two \pseudo\ spin subbands. The number density of electrons is given by
\begin{equation}\label{eq:numberdensity}
N_\lambda = \int^\infty_{-\infty} \frac{dk}{2\pi} f(E) ,
\end{equation}
where $f(E)$ is the distribution function.  In the following, we consider the limiting cases temperature $T=0$, when $f$ is a step function, and high temperature, when $f$ becomes the Maxwell-Boltzmann distribution
\begin{equation}\label{eq:distribution}
  f(E)=
  \begin{cases}
    \theta(E_F - E), &T=0 \\
    e^{-E/k_B T}, & \text{high}~T.
  \end{cases}
\end{equation}
For $T=0$ and a given Fermi energy $E_F$, the Fermi wave vectors for the
dispersion (\ref{eq:dispersion}) become
\begin{equation}\label{eq:fer-wave-vec-exact}
k_{F,\lambda}^{(\pm)} =  \pm \sqrt{k_F^2 + \frac{\alpha^2}{\mu^2}}- \frac{\lambda \alpha}{\mu} ,
\end{equation}
where $k_F \equiv \sqrt{2E_F/\mu}$.  Assuming small spin-orbit coupling $\left| \alpha / \mu k_F \right|\ll1$, Eq.\ (\ref{eq:fer-wave-vec-exact}) reduces to
\begin{equation}\label{eq:fermivecapprox}
  k_{F,\lambda}^{(\pm)} \approx  \pm k_F-\frac{\lambda \alpha}{\mu}
\end{equation}
so that the number density in equilibrium becomes
\begin{equation}
  N^\mathrm{eq}_\lambda = \frac{1}{2\pi}\left[ k^{(+)}_{F,\lambda}-k^{(-)}_{F,\lambda}\right] \approx \frac{k_F}{\pi} .
\end{equation}
Similarly, at high temperature we define the thermal wave vector $k_T \equiv \sqrt{\pi k_B T / 2\mu}$. Then, assuming $\left|\alpha / \mu k_T \right|\ll 1$, we obtain
\begin{equation}\label{eq:equil-density-hightemp}
N^\mathrm{eq}_\lambda =\frac{k_T}{ \pi} .
\end{equation}

\section{Position and Velocity Operators}
\label{sec:vel-op}

The position and velocity operators in multiband systems such as
those described by the Hamiltonian of Eq.\ (\ref{eq:ham:4x4}) are
important quantities for a wide range of topics, some of which will
be discussed below. It has long been debated \cite{yaf63, noz73,
bae90, eng07a, bi13} whether the position operator in a multiband
system, when reduced to the subspace of a single band, should
acquire a spin-dependent correction that leads to a doubling of the
spin-dependent correction for the corresponding velocity
operator. We review this question for the particular example of the
universal Hamiltonians (\ref{eq:ham:4x4}), (\ref{eq:ham:2x2}), and
(\ref{eq:ham:2x2:1d}).  Based on our findings, we argue that the
spin-dependent correction for the single-band position operator, and
the contribution to the single-band velocity operator arising from
it, should be absent \cite{schrieffer}.

According to the Heisenberg equation of motion for the position
operator $\vek{r}_{4\times 4}$, the velocity operator for the
Hamiltonian $H_{4\times 4}$ becomes \cite{tha92, win07}
\begin{equation}
  \label{eq:vel:4}
  \vek{v}_{4\times 4}
  = \frac{d\vek{r}_{4\times 4}}{dt}
  = \frac{i}{\hbar} [H_{4\times 4}, \vek{r}_{4\times 4}]
  = \frac{\partial H_{4\times 4}}{\hbar \,\partial\vek{k}}
  = \frac{\gamma}{\hbar} \, \vek{\sigma} \rho_x .
\end{equation}
On the other hand, we may define the velocity operator for the
effective $2\times 2$ Hamiltonian $\mathcal{H}_+$ in two different
ways.  In the first approach, we use
\begin{equation}
  \label{eq:vel:2}
  \vek{v}_{2\times 2}
  % = \frac{d\vek{r}_{2\times 2}}{dt}
  = \frac{i}{\hbar} [\mathcal{H}_+, \vek{r}_{2\times 2}]
  = \frac{\partial \mathcal{H}_+}{\hbar \,\partial\vek{k}}
  = \frac{1}{\hbar} \left(\frac{2\gamma^2}{\Delta} \vek{k}
    - \frac{\gamma^2}{\Delta^2} \vekc{E} \times \vek{\sigma} \right) .
\end{equation}
In the second approach, we remember the fact that $\mathcal{H}_+$
was derived via a FW transformation $e^{-S}$ from the
Hamiltonian $H_{4\times 4}$.  Accordingly, we first apply the same
unitary transformation to $\vek{r}_{4\times 4}$, which yields
\begin{subequations}
\begin{align}
  \vek{r}_{4\times 4}^\mathrm{FW}
  \equiv & \; e^{-S} \vek{r}_{4\times 4} \, e^S , \\
  = & \; \vek{r}_{4\times 4}
    + \frac{\gamma^2}{\Delta^2} \kk \times \vek{\sigma} \rho_0
    - \frac{\gamma}{\Delta} \vek{\sigma} \rho_y , \\
  = & \; \tvek[cc]{ \vek{r}_{2\times 2}^\mathrm{FW}
    & \frac{i\gamma}{\Delta}\, \vek{\sigma}\\[0.2cm]
    \frac{-i\gamma}{\Delta}\, \vek{\sigma} & \vek{r}_{2\times 2}^\mathrm{FW}} .
\end{align}
\end{subequations}
Unlike the transformed Hamiltonian $H_{4\times 4}^\mathrm{FW}$, the
transformed FW position operator $\vek{r}_{4\times 4}^\mathrm{FW}$
does not acquire a block-diagonal form.  Ignoring the off-diagonal part that
couples the subspaces of positive and negative energies, we get the
following modified FW position operator for the subspace of
$\mathcal{H}_+$
\begin{equation}
  \vek{r}_{2\times 2}^\mathrm{FW}
  = \vek{r}_{2\times 2}
    + \frac{\gamma^2}{\Delta^2} \kk \times \vek{\sigma} .
\end{equation}
Alluding to Ref.~\onlinecite{yaf63}, the spin-dependent part of $\vek{r}_{2\times 2}^\mathrm{FW}$ has sometimes been called the Yafet term  \cite{eng07a}.  The modified FW position operator yields the velocity operator
\begin{equation}
  \label{eq:vel:2:mod}
  \vek{v}_{2\times 2}^\mathrm{FW}
  = \frac{i}{\hbar} [\mathcal{H}_+, \vek{r}_{2\times 2}^\mathrm{FW}]
  = \frac{1}{\hbar} \left(\frac{2\gamma^2}{\Delta} \vek{k}
    - \frac{2\gamma^2}{\Delta^2} \vekc{E} \times \vek{\sigma} \right) .
\end{equation}
The same result (\ref{eq:vel:2:mod}) is obtained if the FW
transformation is applied to the velocity operator
$\vek{v}_{4\times 4}$, which yields \cite{velocity}
\begin{subequations}
  \label{eq:vel:4:mod}
  \begin{align}
  \vek{v}_{4\times 4}^\mathrm{FW}
  \equiv & \; e^{-S} \vek{v}_{4\times 4} \, e^S , \\
  = & \; \frac{1}{\hbar} \left[\frac{2\gamma^2}{\Delta} \vek{k}
    - \frac{2\gamma^2}{\Delta^2} \vekc{E} \times \vek{\sigma} \rho_0 
    + \gamma \vek{\sigma} \rho_x
    + \frac{\gamma^3}{\hbar\Delta^2} \vek{k} (\vek{k} \cdot \vek{\sigma}) \rho_x
    \right] , \\
    =& \; \tvek[cc]{ \vek{v}_{2\times 2}^\mathrm{FW} 
    & \frac{\gamma}{\hbar}\, \vek{\sigma}
    + \frac{\gamma^3}{\hbar\Delta^2} \vek{k} (\vek{k} \cdot \vek{\sigma})
    \\[0.2cm]
    \frac{\gamma}{\hbar}\, \vek{\sigma}
    + \frac{\gamma^3}{\hbar\Delta^2} \vek{k} (\vek{k} \cdot \vek{\sigma})
    & \vek{v}_{2\times 2}^\mathrm{FW} } . \end{align}
\end{subequations}
Ignoring the off-diagonal blocks we reproduce Eq.\
(\ref{eq:vel:2:mod}).  Comparing Eqs.\ (\ref{eq:vel:2}) and
(\ref{eq:vel:2:mod}) we see that the spin-dependent parts
characterizing $\vek{v}_{2\times 2}$ and
$\vek{v}_{2\times 2}^\mathrm{FW}$ differ by a factor of two.  The
significance of this factor of two has been debated in the past
\cite{yaf63, noz73, bae90, eng07a, bi13}.

\begin{figure}
  \includegraphics[width=0.45\columnwidth]{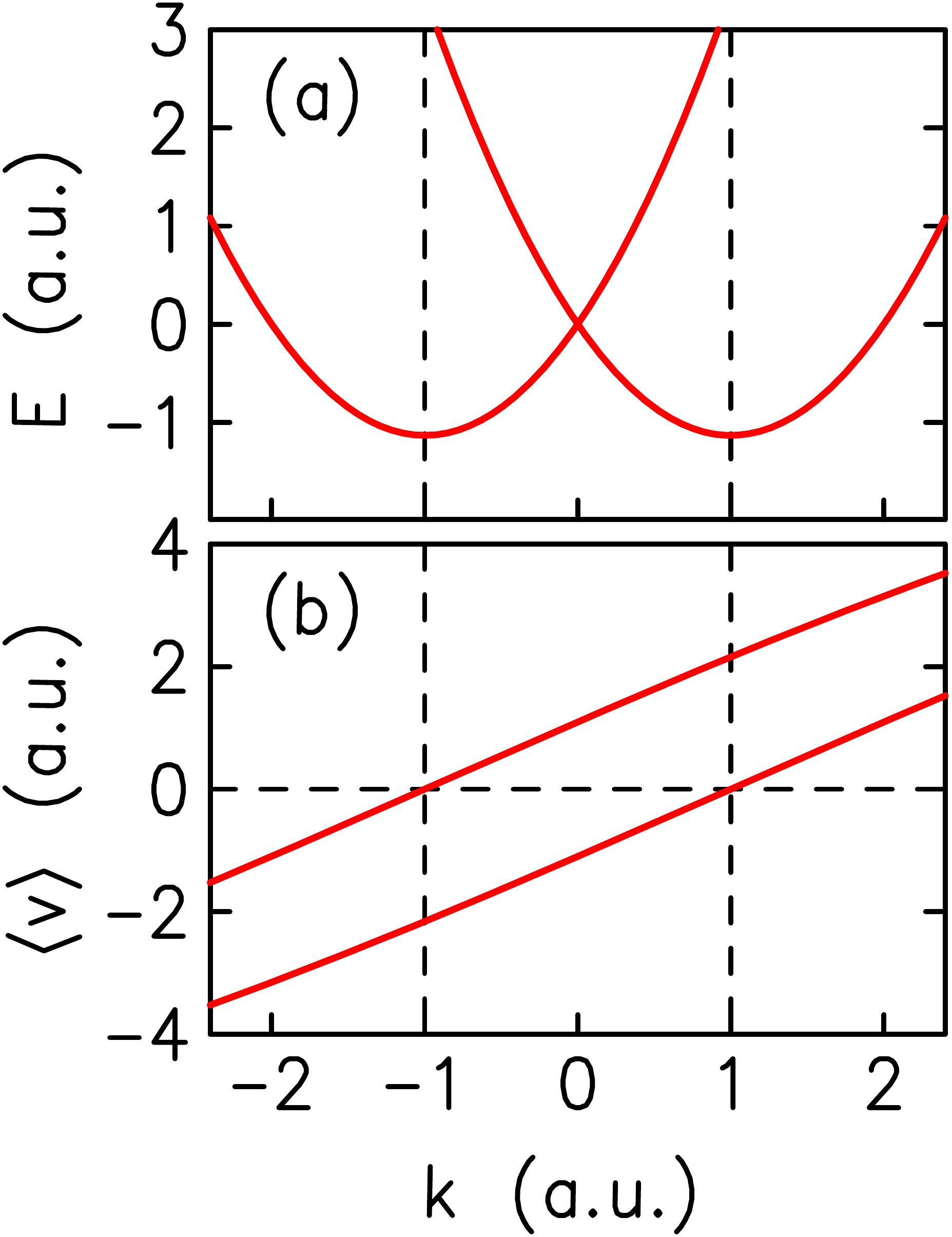}
  \caption{\label{fig:velocity} (a) Spin-split dispersion and (b)
  expectation value of the velocity $v_{4\times 4}$ for a quantum
  wire described by the $4\times 4$ Hamiltonian (\ref{eq:ham:4x4})
  augmented by an inversion-asymmetric mass confinement in the $y$
  direction.  The numerical results shown here were obtained using
  the quadrature method described in Ref.~\onlinecite{win93a}.}
\end{figure}

Focusing on quasi-1D wires discussed here, the Hellmann-Feynman theorem
applied to Eq.\ (\ref{eq:vel:4}) yields
\begin{equation}
  \label{eq:hellmann:4}
  \braket{v_{4\times 4}} (k)
  = \left\langle \frac{\partial H_{4\times 4}}{\hbar \,\partial k}
    \right\rangle
  = \frac{\partial E_{\lambda, 4\times 4}}{\hbar \,\partial k} ,
\end{equation}
where the expectation value is taken for the four-component
wire eigenstates of $H_{4\times 4} + V(y) \sigma_0 \rho_z$
with eigenvalues $E_{\lambda, 4\times 4} (k)$, and $V(y)$ is the
confining potential of the quantum wire.  Accordingly, the
expectation value $\braket{v_{4\times 4}}$ as a function of $k$
changes sign at the extrema of the dispersion
$E_{\lambda, 4\times 4} (k)$.  This is illustrated in
Fig.~\ref{fig:velocity} for a quasi-1D wire realized via
an inversion-asymmetric confinement $V(y)$.  On the other hand,
the $2\times 2$ model (\ref{eq:ham:2x2:1d}) yields
\begin{subequations}
\label{eq:vel:2:1d}
\begin{align} 
  v = & [\mathcal{H},x] =
  \frac{\partial \mathcal{H}}{\hbar \,\partial k}
    = \frac{1}{\hbar} (\mu k + \alpha \sigma_z), \\
  v^\mathrm{FW} = &  [\mathcal{H},x^\mathrm{FW}] 
  = \frac{1}{\hbar} (\mu k + 2\alpha \sigma_z) .
\end{align}
\end{subequations}
Thus an important difference between $v$ and $v^\mathrm{FW}$ lies
in the fact that the extrema of the dispersion (\ref{eq:dispersion})
at $k_{0,\lambda} \equiv - \lambda \alpha/\mu$ are characterized by
$\braket{v} (k_{0,\lambda}) = 0$, but $\braket{v^\mathrm{FW}} (k_{0,\lambda}) \ne 0$.

The FW transformation $e^{-S}$ is set up with the goal to yield the Hamiltonian for the effective one-band model that faithfully reproduces the features of the full two-band theory at low energies. Assuming that the relation (\ref{eq:hellmann:4}) between velocity and dispersion should also hold for a system described by the one-band Hamiltonian (\ref{eq:ham:2x2:1d}), the Hellmann-Feynman theorem implies
\begin{equation}
  \label{eq:hellmann:2}
  \braket{v} (k)
  = \frac{\partial E_\lambda}{\hbar \,\partial k}
  = \left\langle \frac{\partial \mathcal{H}}{\hbar \,\partial k}
    \right\rangle ,
\end{equation}
where the expectation value is now taken for the two-component
eigenstates of $\mathcal{H}$ with eigenvalues $E_\lambda (k)$.
Comparing Eq.\ (\ref{eq:hellmann:2}) with Eq.\ (\ref{eq:vel:2:1d}),
we conclude that only $v$ is consistent with Eq.\
(\ref{eq:hellmann:4}), i.e., for both the two-band Hamiltonian
(\ref{eq:ham:4x4}) and the one-band Hamiltonians (\ref{eq:ham:2x2})
and (\ref{eq:ham:2x2:1d}) the position operator should not include a
spin-dependent term.  We speculate that the discrepancy between the
predictions from $v^\mathrm{FW}$ and Eq.\ (\ref{eq:hellmann:4}) may
be due to the fact that the approximate FW transformation $e^{-S}$
used to derive $\vek{v}_{2\times 2}^\mathrm{FW}$ from
$\vek{v}_{4\times 4}$ (and $\vek{r}_{2\times 2}^\mathrm{FW}$ from
$\vek{r}_{4\times 4}$) is applied to \emph{operators} so that it is
generally difficult to estimate the magnitude of the omitted terms
\cite{tha92}.  Note that the terms omitted when going from Eq.\
(\ref{eq:vel:4:mod}) to Eq.\ (\ref{eq:vel:2:mod}) include the
original Dirac velocity operator (\ref{eq:vel:4}) whose matrix
elements may contribute substantially to expectation values.

\section{Universal (Pseudo-) Spin Dynamics in Quasi-1D Quantum Wires}

In the following we discuss a few examples for the universal
(pseudo-) spin dynamics in quasi-1D quantum wires emerging from the
Hamiltonian (\ref{eq:ham:2x2:1d}).

\subsection{Equilibrium Spin Currents}
\label{sec:eq-spin-cur}

\begin{figure}
    \includegraphics[width=0.95\columnwidth]{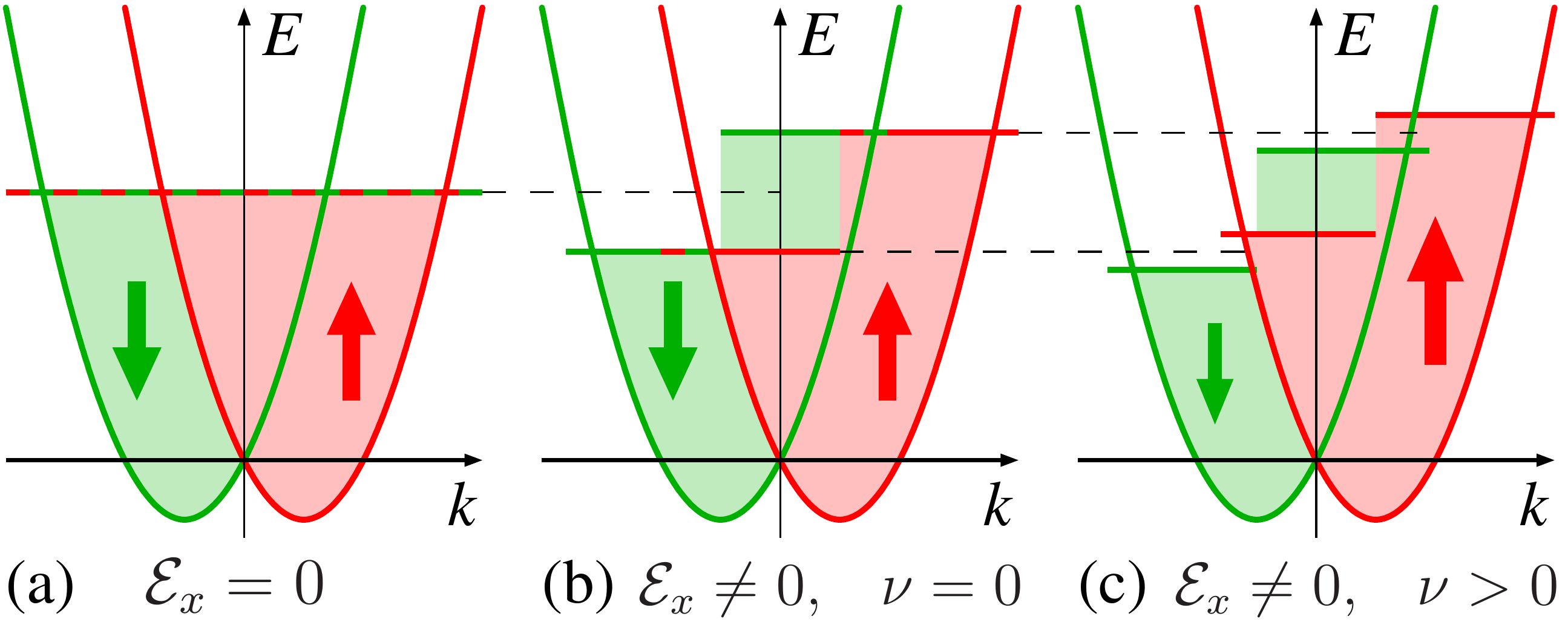}
\caption{Qualitative sketch of the dispersion $E(k)$ of a quantum wire with Rashba-like SO coupling at $T=0$ (a) in thermal equilibrium with electric field $\mathcal{E}_x = 0$, (b) in the presence of a driving electric field $\mathcal{E}_x > 0$ along the wire and a dissipative regime with $\nu = 0$, and (c) $\mathcal{E}_x > 0$ with $\nu > 0$.  Horizontal colored lines indicate the quasi Fermi levels for left and right movers in the two spin subbands. In (a) and (b) the net spin polarization and the spin current are exactly zero. In (c) the driving electric field results in a steady state with a net spin polarization (Edelstein effect).} \label{fig:spinsplit}
\end{figure}

Using Rashba's definition\cite{ras03}, the spin current operator for the velocity operator $v$ in Eq.\ (\ref{eq:vel:2:1d}) becomes
\begin{equation}\label{eq:vel-op}
  j_s \equiv \lbrace v,\sigma_z \rbrace =\frac{1}{\hbar}\left( \mu k \sigma_z + \alpha \right) ,
\end{equation}
where $\lbrace A,B \rbrace \equiv \frac{1}{2}\ (AB + BA)$.
The total average spin current is obtained using
\begin{equation}\label{eq:av-spin-cur}
I_s = \braket{j_s} =
\braket{\lbrace v,\sigma_z \rbrace} = \sum_{\lambda = \pm} \int \frac{dk}{2 \pi} \braket{ \lambda | \lbrace v,\sigma_z \rbrace | \lambda }.
\end{equation}
Equation (\ref{eq:dispersion}) shows that the parabolic dispersion curves for the two spin subbands are centered about $-\lambda \alpha/\mu$ [Fig.\ \ref{fig:spinsplit}(a)].  As expected \cite{mal03, kis05}, this yields $\braket{j_s} = 0$. This holds for both $T=0$ and high temperatures.

For comparison, we note that the spin current for the modified FW velocity operator becomes
\begin{equation}
 j_s^\mathrm{FW} = \lbrace v^\mathrm{FW},\sigma_z \rbrace =
 \frack{1}{\hbar}\left( \mu k \sigma_z + 2\alpha \right) = j_s + \alpha/\hbar .
\end{equation}
For both $T=0$ and high temperatures, the total equilibrium spin current (\ref{eq:av-spin-cur}) then becomes $(\alpha /\hbar) N^\mathrm{eq}$, where $N^\mathrm{eq} = N^\mathrm{eq}_+ + N^\mathrm{eq}_-$ is the total electron density in equilibrium given by Eq.\ (\ref{eq:numberdensity}).

\subsection{Edelstein Effect}
\label{sec:edelstein}

We consider a driving electric field $\mathcal{E}_x$ along the direction of the wire.  In a dissipative regime, using a Drude model \cite{cha90}, the distribution is then shifted from $f(k)$ to $f[k+k_\nu^d(k)]$, where $k_\nu^d(k) = e \mathcal{E}_x \tau_\nu(k)/\hbar$. This causes a net motion of electrons depicted in Fig.\ \ref{fig:spinsplit}(b), where more spin-up states contribute to the charge current than spin-down states. This phenomenon resulting in a net spin polarization is often called the Edelstein effect \cite{ede90}, see also Refs.~\onlinecite{ivc78, aron91}.  Here we evaluate the Edelstein effect for a quasi-1D quantum wire characterized by the generic Hamiltonian (\ref{eq:ham:2x2:1d}).  Recently, a valley Edelstein effect in 2D systems has been discussed in Ref.~\onlinecite{tag17}.

We express the scattering time $\tau_\nu$ as a power law \cite{kai04}
\begin{equation}
\tau_\nu (k) = \zeta_v k^{2\nu}, \quad \nu = 0,1,2,
\end{equation}
where $\zeta_\nu$ is a proportionality constant. The parameter $\nu$
depends on the scattering mechanism \cite{aron91, kai04}.
Scattering by, e.g., acoustic and optical phonons and screened
ionized impurities corresponds to the case $\nu=0$. The case $\nu=1$
pertains to piezoelectric scattering by acoustic phonons or
scattering by polar optical phonons.  Scattering by weakly screened
ionized impurities belongs to the case $\nu=2$.  We define the net
spin polarization as
\begin{equation}
\mathcal{P} = \frac{N_+-N_-}{N_++N_-} ,
\end{equation}
where $N_\lambda$ is the number density for each spin subband~$\lambda$.  As expected, in thermal equilibrium ($\mathcal{E}_x = 0$) we have $\mathcal{P} = 0$ [Fig.~\ref{fig:spinsplit}(a)].

\subsubsection{Zero Temperature}
\label{sec:edelstein-zerotemp}

In the zero temperature case with driving electric field, the distribution becomes a shifted step function. For a weak electric field $\mathcal{E}_x$, we can assume that $| k^d_\nu(k_F) / k_F | \ll 1$ so that the extremal wave vectors for the occupied states are approximately
\begin{equation}
k^{(\pm)} =k_{F,\lambda}^{(\pm)}-k_\nu^d(k_{F,\lambda}^{(\pm)}).
\end{equation}
This yields quasi Fermi levels for left and right movers in the two spin subbands
\begin{equation}
  E_{F,\lambda}^{(\pm)} = E_F + \frac{ek_F^{2\nu}
  \mathcal{E}_x\zeta_\nu}{\hbar} 
  \left(\mp \mu k_F + 2 \lambda \nu \alpha \right),
\end{equation}
where we assumed small SO coupling $\left| \alpha / \mu k_F \right|\ll1$.
Thus we have two contributions for the $\mathcal{E}_x$-dependent corrections to the quasi-Fermi levels $E_{F,\lambda}^{(\pm)}$: A spin-independent correction $\propto \mp k_F$ raises $E_F$ for the right movers and lowers $E_F$ for the left movers.  For $\nu > 0$, the spin-dependent correction $\propto \lambda \nu \alpha$ is positive for both right- and left movers in one spin subband, and it is negative for the other spin subband, corresponding to a transfer of electrons from one spin subband to the other.

The Edelstein effect also becomes explicit by looking at the number densities $N_\lambda$ as a function of driving field $\mathcal{E}_x$.  For small SO coupling $\left| \alpha / \mu k_F \right|\ll1$ we get
\begin{equation}\label{eq:nonequilibriumdensitygeneral}
N_{\lambda} =N_\lambda^\mathrm{eq}\left[ 1 + \lambda \frac{2\nu \alpha k_\nu^d(k_F)}{\mu k_F^2} \right]
\end{equation}
and the polarization for any $\nu$ is
\begin{equation}\label{eq:polarizationlowtemp}
\mathcal{P}=  2 \nu \frac{\alpha k_\nu^d(k_F)}{\mu k_F^2}.
\end{equation}

The trivial case $\nu = 0$ is one where the scattering time $\tau$ is a constant independent of the wave vector $k$. As can be seen in Eq.\ (\ref{eq:nonequilibriumdensitygeneral}), the number densities for spin up and spin down subbands are equal, which means that there is no net spin polarization, $\mathcal{P} = 0$. This also means that a constant shift $k_0^d$ in the electron distribution does not affect the number density in each subband. On the other hand, an unequal shift in the wave vector $k^{(+)}$ and $k^{(-)}$ results in an imbalance among spin-up and spin-down states, yielding a nonzero spin polarization.

We define the average scattering time $\bar{\tau}_\nu$ for any $\nu$ as \citep{aron91}
\begin{equation}\label{eq:taup}
\bar{\tau}_\nu = \frac{\braket{\tau_\nu E}}{\braket{E}} ,
\end{equation}
where for a function $\phi$, the average $\braket{\phi}$ is defined as
\begin{equation}
\label{eq:av-dev}
\braket{\phi} = \frac{\int_{-\infty}^{+\infty} \phi(k) f(k) dk}{\int_{-\infty}^{+\infty}f(k)dk}.
\end{equation}
We also define $\bar{k}^d_\nu \equiv e \mathcal{E}_x \bar{\tau}_\nu/\hbar$ so that Eq.\ (\ref{eq:polarizationlowtemp}) can then be written generally as \citep{aron91}
\begin{equation}\label{eq:gen-pol}
\mathcal{P}=Q\frac{\alpha \bar{k}^d_1}{\braket{E}} ,
\end{equation}
where $Q$ is a dimensionless number that depends on $\nu$. Its
value for the different limiting temperatures are listed in Table \ref{table:nconst}.

\subsubsection{High Temperature}
\label{edelstein-hightemp}

Similar to the case $T=0$,
we can derive the number density for each spin subband, and hence the spin polarization at high temperature by shifting the Maxwell-Boltzmann distribution in Eq.\ (\ref{eq:distribution}) by $k_\nu^d(k)$.  Again we assume  small spin-orbit coupling $\left| \alpha / \mu k_T \right| \ll 1$ and weak electric fields $| k^d_\nu (k_T) / k_T | \ll 1$. The number density then simplifies to
\begin{equation}\label{eq:densityapprox}
N_\lambda =N^\mathrm{eq}_\lambda\left[1 + \lambda\frac{\pi k^d_\nu (k_T)}{k_T} \left( \frac{4}{\pi} \right)^\nu  \frac{\nu(2 \nu  - 1)!!}{2^\nu}  \frac{ \alpha}{\mu k_T} \right] ,
\end{equation}
which yields the polarization
\begin{equation}\label{eq:polarizationhightemp}
\mathcal{P} = \frac{\pi \alpha k^d_\nu (k_T)}{\mu k_T^2} \left( \frac{4}{\pi} \right)^\nu  \frac{\nu(2 \nu  - 1)!!}{2^\nu} .
\end{equation}

Similar to zero temperature, the case $\nu = 0$ gives the same number densities for both spin subbands [see Eq.\ (\ref{eq:densityapprox})] so that $N_+-N_- = 0$ and hence $\mathcal{P} = 0$. For the cases $\nu = 1,2$, we obtain a polarization of the form (\ref{eq:gen-pol}) with values of $Q$ listed in Table \ref{table:nconst}.

\begin{table}
\caption{Numerical value for $Q$ at the limiting temperatures for $\nu=0,1,2$.}\label{table:nconst}
\tabcolsep 2em
\begin{tabular}{CCc} \hline\hline
Q & T=0 & high $T$\\
\hline
\nu=0 &  0   & 0 \\
\nu=1 &  5/9 & 1/3 \\
\nu=2 & 14/9 & 2/5 \\ \hline\hline
\end{tabular}
\end{table}

\subsection{Adiabatic (Pseudo-) Spin Pumping}
\label{sec:spinpump}

\begin{figure}
\includegraphics[width=0.7\columnwidth]{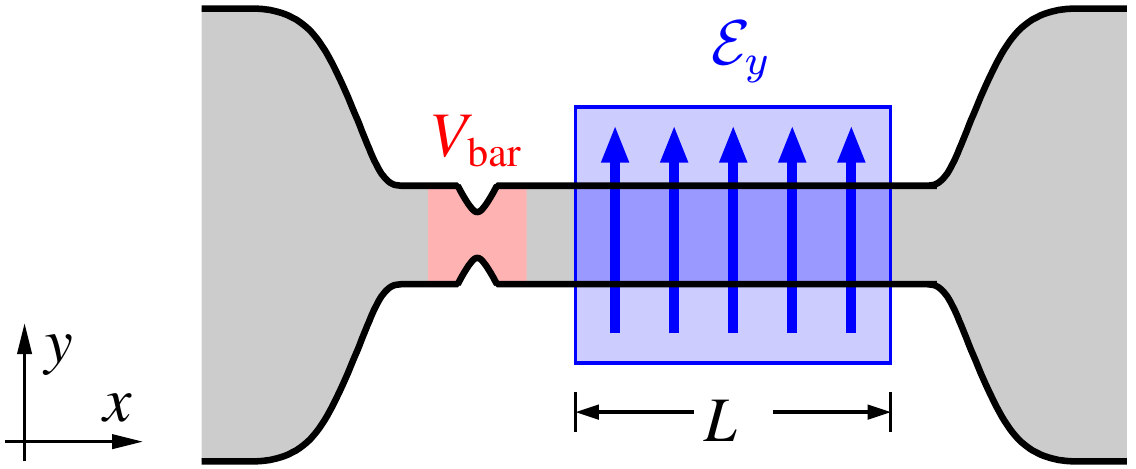}
\caption{Schematic diagram for an adiabatic \pseudo\ spin pump \cite{gov03}. A potential barrier $V_\mathrm{bar}$ is present at the left end of the wire. A perpendicular electric field $\mathcal{E}_y$ is applied in the blue shaded region of length $L$ in order to tune the coupling coefficient $\alpha \propto \mathcal{E}_y$.}\label{fig:valleypumping}
\end{figure}

In quantum wires obeying the Hamiltonian (\ref{eq:ham:2x2:1d}), dc
spin currents can be generated via parametric pumping, \cite{bro98, gov03} where one varies periodically a potential barrier $V_\mathrm{bar}$ in the wire and the electric field perpendicular to it  (Fig.~\ref{fig:valleypumping}).  Here the total Hamiltonian becomes
\begin{equation}
H = \mathcal{H} + V_\mathrm{bar}.
\end{equation}
Assuming that the potential barrier is a $\delta$ potential, $V_\mathrm{bar} = V \delta(x)$, the spin-$\lambda$ particle current is derived using the parametric integral \citep{bro98}
\begin{equation}
I_\lambda = \frac{\omega}{2 \pi^2}\int_A d V d \alpha 
\Im \left( \frac{\partial r_\lambda^\ast}{\partial V}
  \frac{\partial r_\lambda}{\partial \alpha}
+ \frac{\partial t_\lambda^\ast}{\partial V}
  \frac{\partial t_\lambda}{\partial \alpha} \right),
\end{equation}
where $r_\lambda$ and $t_\lambda$ are the reflection and transmission coefficients, respectively, for particles with \pseudo\ spin $\lambda$, and the integral is over the area $A$ enclosed by the path in the parameter space $(V,\alpha)$.  For a sinusoidal pumping cycle with $V = V_0 + \Delta V \sin (\omega t)$ and $\alpha = \alpha_0 + \Delta \alpha \sin(\omega t - \phi)$ and assuming the weak-pumping limit $\Delta V \ll V_0$ and $\Delta \alpha \ll \alpha_0$, the total spin current $I_s = I_+ - I_-$ becomes \cite{gov03}
\begin{subequations}
\label{eq:spincur:pump}
\begin{align}
I_s = & \frac{\omega}{\pi} \sin(\phi) \Delta V \Delta\alpha
\frac{\mu k^2_F L V_0}{\left(\mu^2 k_F^2 +V_0^2\right)^2} \\
\equiv & I_s^{(0)} \, \omega \sin(\phi)
\frac{\Delta V}{V_0} \frac{\Delta\alpha}{\alpha_0}
\end{align}
with dimensionless prefactor
\begin{equation}
  I_s^{(0)} = \frac{1}{\pi}
  \frac{\mu k^2_F L V_0^2 \alpha_0}{\left(\mu^2 k_F^2 +V_0^2\right)^2} ,
\end{equation}
\end{subequations}
where $L$ is the length of the region of the wire and $\alpha$ is
modulated by tuning the field $\mathcal{E}_y$ (Fig.~\ref{fig:valleypumping}). 
The maximum spin current is achieved when $V_0 \simeq \mu k_F$, which corresponds to a barrier $V_0 \delta(x)$ with transmission probability $\mathcal{T} \simeq 1/2$. Also, this corresponds to a maximum~$I_s^{(0)}$
\begin{equation}
  I_s^{(0),\mathrm{max}} \simeq  \frac{eL\mathcal{E}_y}{8\pi\Delta} ,
\end{equation}
implying that only the energy gap $\Delta$ characterizes the materials' effectiveness for operating a (valley) spin pump.  In particular,  $I_s^{(0),\mathrm{max}}$ is independent of the Fermi wave vector or density.  For a length $L \simeq 10~\mu\mathrm{m}$ and lateral electric field $\mathcal{E}_y \simeq 1~\mathrm{kV/cm}$, we have $I_s^{(0),\mathrm{max}} \simeq 0.1$ for InAs, InSb and graphene and $I_s^{(0),\mathrm{max}} \simeq 0.01$ for TMDCs. For frequencies $\omega \simeq 10^4 - 10^5~\mathrm{s}^{-1}$ and $\frac{\Delta V}{V_0} \frac{\Delta \alpha}{\alpha_0} \simeq 0.1$, the spin current becomes $I_s \simeq 100~\mathrm{s}^{-1}$,
which is comparable to the charge currents in single-electron transistors \cite{sch04d}.

Equation (\ref{eq:spincur:pump}) represents a scheme to generate spin currents in a quantum wire that relies on adiabatic pumping \cite{bro98, gov03}.  An alternative scheme operating in a dissipative regime and likewise applicable to the different realizations of the Hamiltonian (\ref{eq:ham:2x2:1d}) was discussed in Ref.~\onlinecite{mal03}.

\section{Conclusion}
\label{sec:conclusion}

In this paper we showed that a general effective Hamiltonian can be formulated to describe spin-dependent phenomena in low-dimensional systems that is realized in a range of different materials. The spin $\sigma_i$ appearing in Hamiltonian (\ref{eq:ham:4x4}) corresponds to the real spin in Dirac or Kane systems, whereas it represents the valley pseudospin in graphene and a combination of valley and real spins in TMDCs.  The universal nature of the Hamiltonian (\ref{eq:ham:4x4}) implies that the spin dynamics present in one of these system exists similarly in the other systems realizing Hamiltonian (\ref{eq:ham:4x4}).  On a qualitative level, the universality of the dynamics is not affected by perturbations such as (pseudo) spin relaxation, while specific numbers for the various systems are certainly different as illustrated by the material parameters listed in Table~\ref{table:parameter}.  Projecting the two-band Hamiltonian (\ref{eq:ham:4x4}) on the conduction or valence band yields the effective single-band Hamiltonian (\ref{eq:ham:2x2}).  In order to describe quasi-1D systems the latter can be further simplified, yielding Eq.\ (\ref{eq:ham:2x2:1d}).

A comparison between the effective one-band Hamiltonian (\ref{eq:ham:2x2:1d}) and the more complete two-band Hamiltonian (\ref{eq:ham:4x4}) allowed us to identify the correct form of the \pseudo\ spin-dependent velocity operator to be used in a single-band theory.  We have shown that equilibrium \pseudo\ spin currents vanish in quasi-1D systems governed by the Hamiltonian (\ref{eq:ham:2x2:1d}). We have also studied the Edelstein effect for quantum wires, whereby in a dissipative regime a driving electric field induces a (pseudo-) spin polarization. This effect vanishes in quasi-1D wires where the scattering time $\tau$ is independent of the wave vector $k$. For $\tau \propto k^{2\nu}$, $\nu \ge 0$, the spin polarization is given by Eq.\ (\ref{eq:gen-pol}). Lastly, we considered adiabatic spin pumping in quasi-1D wires. The induced spin current can be optimized by choosing a critical barrier strength $V_0 \simeq \mu k_F$. For realistic values of system parameters, the maximum spin current is $I_s \simeq 100 s^{-1}$. We have only presented here a limited number of examples illustrating the universal \pseudo\ spin dynamics in low-dimensional systems emerging from the generic Hamiltonian (\ref{eq:ham:4x4}).  More examples can be identified.

\begin{acknowledgments}
  We appreciate stimulating discussions with G. Burkard, D. Culcer,
  M. Governale, A. Korm\'anyos, Q. Niu, E. Rashba, and D. Xiao.  RW
  appreciates the hospitality of Victoria University of Wellington,
  where part of this work was performed.  This work was supported by
  the NSF under Grant No.\ DMR-1310199.  Work at Argonne was
  supported by DOE BES under Contract No.\ DE-AC02-06CH11357.
\end{acknowledgments}

\end{document}